# A Wearable EEG System for Closed-Loop Neuromodulation of High-Frequency Sleep-Related Oscillations

**Authors:** Scott Bressler[1], Ryan Neely[1]*, Heather Read[1,2], Ryan Yost[1], David Wang[1]

**Affiliations**
[1]Elemind Technologies, Inc.
[2]Department of Psychological Sciences - Behavioral Neuroscience Division, University of Connecticut
*Corresponding author. Email: ryan.neely@elemindtech.com

**Abstract**

*Objective.* In healthy sleepers, cortical alpha oscillations are present during the transition from wakefulness to sleep, and then dissipate at sleep onset. For individuals with insomnia, alpha power is elevated during the wake-sleep transition and can persist throughout the night. Neuromodulation techniques using phase-locked auditory stimulation to augment or suppress oscillations have been put forth as alternatives to drugs for improving sleep quality. This approach has been applied to slow oscillations present during deep sleep, but due to technical limitations in signal readout it has not been tested on faster frequency alpha oscillations. *Approach.* Here we examine the feasibility of using an endpoint-corrected version of the Hilbert Transform (ecHT) algorithm implemented efficiently on-device to measure alpha phase and deliver phase-locked stimulation in the form of pink noise sound bursts to modulate ongoing alpha oscillations and promote healthy sleep initiation. First, the ecHT algorithm is implemented on a tabletop electroencephalogram (EEG) device and used to measure the timing of the auditory evoked response and its delivery at precise phases of the alpha oscillation. Secondly, a pilot at-home study tests feasibility to use a headband wearable version of the neuromodulation device for real-time phase-locked stimulation in the alpha (8-12 Hz) frequency range. *Main Results.* Auditory stimulation was delivered at the intended phases of the alpha oscillation with high precision, and alpha oscillations were affected differently by stimuli delivered at opposing phases.Our wearable system was capable of measuring sleep micro- and macro-events present in the EEG that were appropriate for clinical sleep scoring during the at-home study. Moreover, sleep onset latencies were reduced for a subset of subjects displaying sleep onset insomnia symptoms in the stimulation condition. *Significance.* This study demonstrates the feasibility of closed-loop real-time tracking and neuromodulation of alpha oscillations using a wearable EEG device. Preliminary results suggest that this approach could be used to accelerate sleep initiation in individuals with objective insomnia symptoms.

## Introduction

A significant fraction of an individual's lifetime is occupied by sleep. The physiological role of sleep is not fully understood, but may encompass several processes necessary for life, including cellular and genomic maintenance, removal of waste products, memory consolidation, and synthesis of essential molecules (Vyazovskiy, 2015). Despite its importance, the 12-month incidence of insomnia for adults in the United States, defined as repeated issues with sleep quality that result in daytime impairment, is 27.3% with an associated loss in quality-adjusted life-years (QUALYs) of 5.6 million (Olfson et al., 2018). A number of pharmacological interventions exist to address insomnia, including benzodiazepines, antidepressants, and antihistamines. However, many such drugs are associated with a significant risk of harm relative to placebo and can be habit-forming (Buscemi et al., 2007). An additional barrier to the development of effective treatments for sleep disorders is the difficulty of objectively measuring sleep quality and quantity. Polysomnography is considered the "gold standard" for the diagnosis of several forms of disordered sleep, but requires experienced operators and often subjects must commit to a multi-night stay in a sleep laboratory. Miniaturization, wireless systems, ambulatory or "home testing", and "big data" approaches have been identified as key features of new technologies to advance the field of sleep medicine (Hirshkowitz, 2016).

To address both the need for effective sleep treatments and new methods for quantifying sleep efficiently, we have developed the Elemind Neuromodulation device (ENMod), an ambulatory wearable electroencephalography (EEG)-based neuromodulation system capable of measuring and recording neural activity associated with sleep as well as the phase of neural oscillations in real-time. This system is able to deliver auditory stimulation locked to arbitrary phases of oscillatory activity, a method which has previously been shown to influence ongoing cortical oscillations (Ngo et al., 2015; Papalambros et al., 2017). Existing approaches have demonstrated sleep-related effects by delivering auditory stimulation locked to the phase of slow-wave oscillations (approximately 0.4-4 Hz),  typically associated with N3 ("Slow Wave") sleep (Debellemaniere et al., 2018; Papalambros et al., 2017).  However, these approaches do not address sleep onset latency (SOL), which is a specific outcome measure identified by the American Association of Sleep Medicine as an important target for insomnia treatments (Edinger et al., 2021; Kang et al., 2018).

The current study explores the feasibility to physiologically track and modulate brain oscillations associated with insomnia utilizing a novel EEG-based neuromodulation device. Elevated EEG power in the alpha band (8-12 Hz) is a physiologic biomarker of insomnia disorder as demonstrated with meta analyses of multiple EEG studies (Riedner et al., 2016; Zhao et al., 2021). Though EEG alpha power is elevated throughout several sleep stages in insomnia disorder, it is most pronounced when subjects are awake and trying to initiate sleep (Riedner et al., 2016).  Hence, elevated alpha may be a sign of cortical hyper-arousal associated with insomnia (Rezaei et al., 2019). Sleep-inducing medications, such as benzodiazepines, have been shown to decrease alpha power at doses that induce sleep and treat insomnia (Berro et al., 2021; Lozano-Soldevilla, 2018). Outside of pharmacological approaches, physiological interventions to disrupt elevated alpha activity are lacking. Though presentation of visual stimuli out-of-phase with alpha rhythms decreases alpha power (Huang et al., 2019);  visual stimulation may not be optimal for sleep health (Phillips et al., 2019). More importantly, phase-estimation algorithms and their associated processing rates historically have limited real-time phase detection accuracy for biological rhythms with higher frequencies, including alpha (Ferster et al., 2022; Grossman et al., 2017; Schreglmann et al., 2021). When implemented as part of an embedded system, our novel end point corrected Hilbert transform (ecHT) algorithm overcomes this limitation, allowing stimulus phase-locking to fast oscillation frequencies. Here, we implement the ecHT algorithms on a wearable EEG neuromodulation device. Using this approach, we track instantaneous alpha phase and deliver audible pink noise sound pulses at  physiologically opposing phases of alpha. We demonstrate the neuromodulatory performance of this technology, including the ability to phase align auditory-evoked response potentials (ERPs)  in the EEG to modulate downstream alpha oscillations.  An in-home ambulatory study confirms this device accurately tracks EEG signals for effective sleep staging and identification of the sleep micro- and macro-events. Finally, this pilot study supports the feasibility of using EEG-based audible phase-locked sound stimulation to reduce sleep onset latency in individuals with sleep initiation problems.

**Methods**

*Subjects and study design for ERP study:*  Initial evoked response potential (ERP) studies were carried out by the Brain Computer Interface Core at the University of Connecticut to measure average auditory ERP latencies with pink-noise sounds and to assess phase-dependent physiological responses. All procedures were approved by the Institutional Review Board at the University of Connecticut. Participants were provided verbal and written descriptions of the experiment procedures and study rationale and they consented to participate in the experiments. Data are reported here as part of a data sharing agreement with Elemind Technologies, Inc. Subjects were recruited locally to take part in a daytime ERP study consisting of three phases: 1) an individual alpha frequency (IAF) measurement, 2) and ERP delay measurement, and 3) a two-phase auditory ERP measurement (see Supplementary Figures and Data). The total experimental time was roughly 50-60 minutes per participant. 22 participants took part in the study, of which 21 were included for analysis. Of those included were 11 males and 10 females, average age 22.75 years (range [18-38]).

*Sleep Study Recruitment, Inclusion and Exclusion Criterion:* Potential study candidates were recruited using advertisements and postings on social media sites (e.g., LinkedIn, Twitter) targeting individuals who regularly report having problems initiating sleep within 30 minutes. Respondents were directed to an online screening survey to determine study eligibility (Table 1). To be included in this study, adults between the ages of 25 and 55 years of age had to meet the following criteria: 1) fluency in English, 2) access to the internet or other cellular data services, and 3) evidence of moderately severe clinical insomnia as determined by an Insomnia Severity Index (ISI) score ≥ 21 (Morin et al., 2011) and a Pittsburgh Sleep Quality Index (PSQI) score > 5 2. We excluded candidates 1) with clinically-confirmed sleep apnea, 2) a current or past history of any neurological or psychiatric disorder, 3) at high risk for generalized anxiety disorder (GAD-7 > 15 3), 4) at moderate to high risk for alcohol abuse disorder (AUDIT-C > 6 4,5), 5) diagnosed with a hearing impairment 6) with a BMI ≥ 33, 7) night shift workers, and 8) pregnant women. Subjects who reported taking antidepressants, stimulants, medication for hypo/hypertension, cannabis or cannabis-derived products, or consuming more than 4 caffeinated beverages per day were also excluded (Table 1). In total, 24 subjects (13 male, 11 female, average age 33.0 ± 6.6 yrs (mean ± SD), median 31 yrs, range 26 to 55 yrs) met our study criteria. All subjects received informed consent following the guidelines outlined by Elemind Technologies' institutional review board, Solutions IRB (Yarnell, AZ).

Table 1: *Inclusion and exclusion criteria used to recruit subjects into the at-home sleep study*

**Inclusion Criteria:**
- Fluency in English
- Access to the internet or other cellular data service
- Evidence of moderately severe clinical insomnia as determined by an Insomnia Severity Index (ISI) score > 21
- PSQI score > 5

**Exclusion Criteria:**
- Clinically-confirmed sleep apnea
- Current or past history of a neurological or psychological disorder
- High risk for generalized anxiety disorder (GAD-7 > 15)
- Moderate to high risk of alcohol abuse disorder (AUDIT-C > 6)
- Diagnosis of hearing impairment
- BMI > 33
- Night shift work
- Pregnant
- Currently taking antidepressants or medication for hypo/hyper-tention
- Use of stimulants, cannabis, or >4 caffeinated beverages/day

*At-home data collection*: All data collected in this study was performed by the subjects themselves while at home. Study hardware and materials were delivered via FedEx or UPS courier service and included our prototype EEG-based neurostimulating headband, an Android smartphone programmed with a companion Elemind app, and a wrist-worn activity tracker (Philips Respironics Actiwatch). All subjects were trained on the use of all study equipment and procedures during a 30-40 minute secure video call. Research personnel were available through email, text, or phone to address any questions or problems.

*Sleep Study Design:* The pilot feasibility sleep study was a single arm, randomized, control, subject-blind, at-home study comparing outcome measures across 3 experimental conditions including 1) *"No Audio"* , 2) *"Alpha Peak Phase-locked Audio",* and 3) *"Alpha Trough Phase-locked Audio."* Both experimental alpha phase-locked sounds were combined with continuous broadband rain sound and delivered for 30 minutes following lights out.

Following the video training call (Wednesday), all subjects were  instructed to wear the headband without any of the data recording or stimulation features enabled for two consecutive nights (Wednesday and Thursday night) to acclimate to the experience of sleeping with the study equipment. Subjects also started wearing their activity trackers and marked their bedtime and wake times throughout the remainder of the study. Beginning on the following week, subjects were randomly assigned one of three different experimental conditions for four consecutive nights (Monday-Thursday night): 1) a control condition with no auditory stimulation,

2) a stimulation condition with pink noise pulses presented during peak alpha phase, and 3) a stimulation condition with pink noise pulses presented during the trough alpha phase. Phase-locked stimulation was presented continuously for 30 minutes for both stimulation conditions (peak and trough).

For each night of the three-week data collection series (Figure 9), subjects were given detailed instructions regarding their nightly routine. Each night, subjects placed the headband on their head before conducting their pre-bedtime routine to ensure the electrodes had enough time to establish contact to their skin. Once in bed, subjects checked the electrode connection strength using the "blink test" and "signal strength" tests (mentioned above). Following a successful signal quality test, subjects were instructed to turn off all other technology devices, shut off the lights, mark their bedtime on the activity watch, start the headband EEG/neurostimulation session, close their eyes, and go to sleep. Researchers stressed the importance of following this exact routine each night in order to ensure consistency of the data collection across nights and subjects.

On the mornings following each data recording session, subjects plugged in their headbands and phones for recharging, uploaded their EEG metadata to Elemind's secure cloud server, and completed a morning survey questionnaire. The morning survey consisted of 11 questions taken from Consensus Sleep Diary (Carney et al., 2012) and additional questions related to their subjective experience wearing our novel neurostimulation headband.

At the conclusion of the three-week study, subjects returned all study equipment back to Elemind and were financially compensated for the participation time.

*Elemind Tabletop Neuromodulation Device Design:* For lab-based studies including measurement of auditory-evoked potentials, we developed a tabletop version of the Elemind Neuromodulation Device. The EEG front end consisted of amplifier stages with 1000x gain, anti-aliasing and powerline filtering, as well as a 24-bit analog to digital converter. Two processing cores were used; the first of which was responsible for sampling the EEG (at 250-500 Hz, selectable) and computation of the endpoint-corrected Hilbert Transform (ecHT) for phase estimation. The other processor/core generated the audio stimulus (>44.1kHz, 16bit stereo) and controlled the audio amplifier. The system architecture for this device is displayed in simplified form in Figure 1.

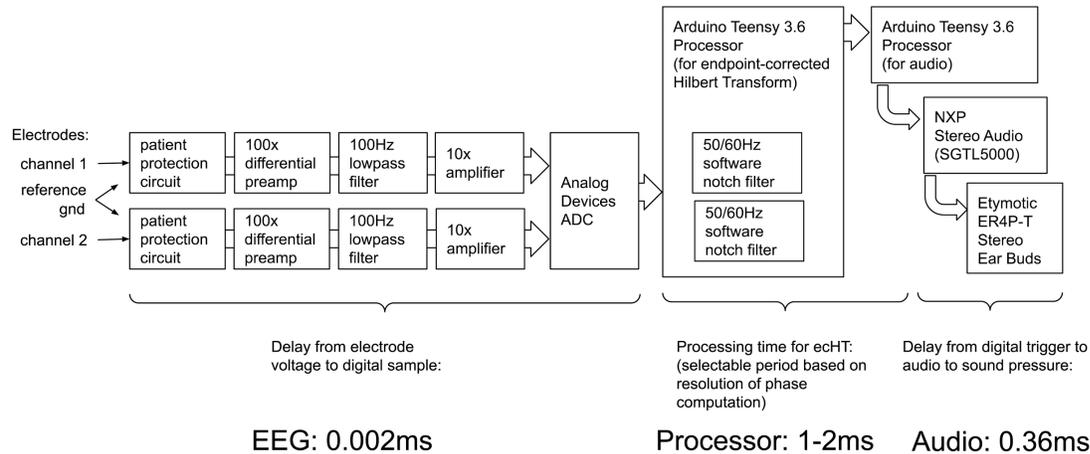

Figure 1: System architecture for the Elemind tabletop ENMod device. The tabletop system featured a 2 channel EEG as input (shown), with expansion ports that include 3x GPIO pins, 2x 12-bit analog outputs, 2x 12-bit analog inputs, and 2x serial ports (not shown). The expansion ports allow researchers to add input and output devices as needed. This figure depicts the major components of the primary neuro-feedback path: the EEG, processing, and audio playback. Below each component is the delay it introduces, which contributes to an end-to-end system delay of ~1.4ms.

*The Elemind Neuromodulation Device (ENMod):* A wearable version of the tabletop research device was produced with dimensions of 70 mm x 45 mm x 20 mm and a weight of 43.9 g, including a single-cell, 1000 mAh capacity rechargeable lithium-polymer battery. The processor unit was designed to attach to a commercially-available fabric headband (Muse-S Gen-2, InteraXon, Toronto, Ontario, Canada) that contained three (3) flexible dry recording electrodes at positions approximating Fp1, Fpz, and Fp2, two linked reference electrodes positioned at the skin just above the ears, and a ground electrode adjacent to Fpz. To eliminate the need for headphones, auditory stimulation was delivered via a bone-conduction driver with a 22 kHz sample rate. Electrode quality check, EEG recording, initiation of experimental conditions and data transfer were all controlled through a smartphone and custom application provided for each subject. Device control with the smartphone was mediated through Bluetooth Low Energy (BLE). Data transfer was possible through Bluetooth or a direct USB-C connection, which was also used for charging. Apart from the app, the user interface on the device was a single RGB LED indicator light, volume up and down buttons, and an "activity" button which was used to perform a hardware reset. The system had other sensors which were not used in this study, including a 3-axis accelerometer, microphone to measure ambient volume, and an ambient light sensor. The firmware and software architecture for this device is shown in Figure 2.

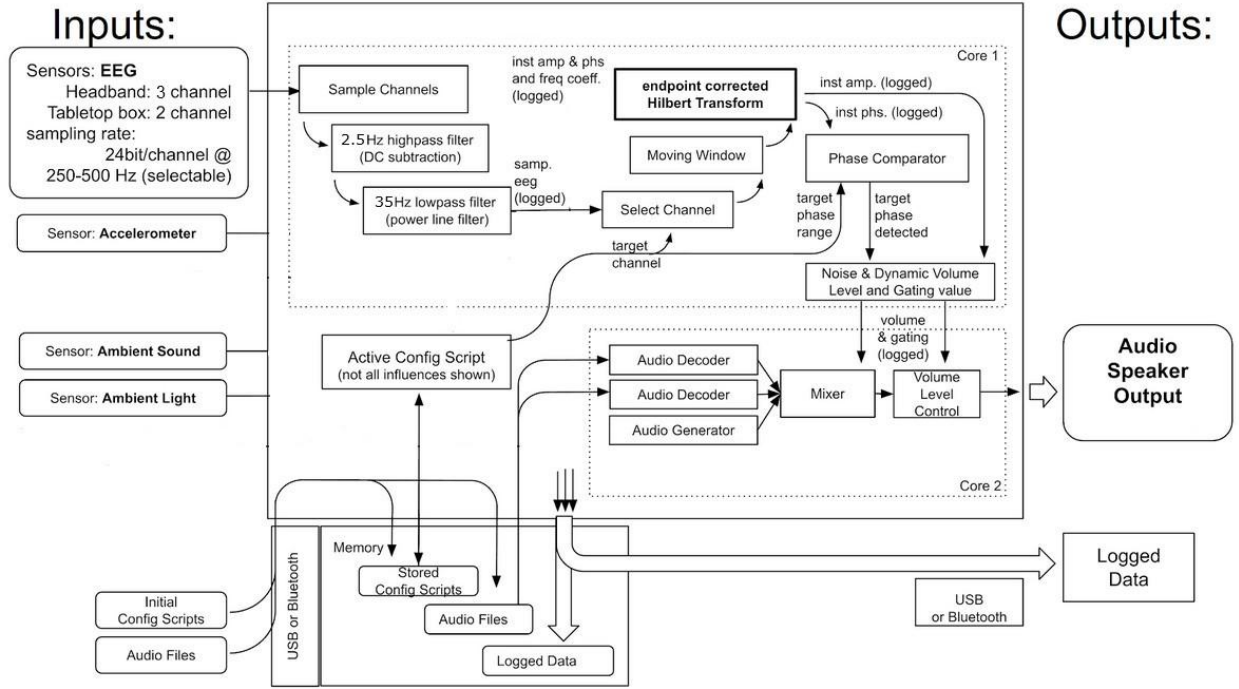

Figure 2: Software architecture for the ambulatory Elemind device. The left side of the diagram shows the inputs: EEG and a variety of ambient sensors. Only EEG sensors were active for this study. Configuration scripts define the parameters used to alter the stimulation paradigm. During the at-home study the scripts represent different study conditions under test (Stim or Control). Core 1 of the processor is responsible for computing when to stimulate, and runs the ecHT algorithm. Core 2 generates the audio and drives the low-latency playback system. Finally, all the sensor data as well as many intermediate computational derivations are logged and can be downloaded or live-streamed for additional (computer, phone, or cloud based) analysis.

*EEG experimental procedures and measures*:  Through a data sharing agreement, we obtained EEG measures of auditory evoked response potential (ERP) latencies from a study conducted at the University of Connecticut. Scalp EEG was recorded with wet electrodes from Fpz with reference at the mastoid (M1) using the tabletop ENMod device (ENModv1, Elemind Technologies, Inc).  First, in a 3 minute recording session with eyes closed, the EEG power spectrum was obtained and 1/f detrended to estimate the individualized alpha frequency (IAF) using standard methods (Klimesch, 1999).  Secondly, in a 15 minute EEG recording session with eyes open, pink-noise sound pulses were played at random phases of ongoing alpha to measure the frontal (Fpz) auditory ERP latency.  Finally, in a 15 minute EEG recording session with eyes closed, subjects were instructed to ignore the sounds while rehearsing multiplication tables for a subsequent verbal test. Pink-noise sound pulses were played to determine whether individually tailored peak and trough phase-locked auditory stimulation had distinct effects on alpha oscillations.

For the in-home study, EEG data collection and real-time phase calculation and sound delivery were  performed on the wearable version of the ecHT devices. EEG signals were sampled at 250 Hz, and then bandpass-filtered with a lowpass cutoff of 35 Hz and a highpass cutoff frequency of 2.5 Hz. The channel with the highest recording quality at a given time was chosen algorithmically to be used for phase estimation. In order to determine the optimal channel, RMS values for the EEG signal were computed for 5-second windows. If the RMS for the current channel dropped below a fixed threshold, the RMS values for the remaining 2 channels were compared, and the channel with the highest value was selected to be active.

*Android App*: The Elemind Android app could be remotely programmed by research staff to schedule the different stimulation conditions. Subjects were asked to start stimulation/data recording on the appropriate nights. It is worth noting that the current design of the prototype headband did not have the ability to measure channel impedances. Subjects instead used the "Electrode Quality Check" smartphone app (Figure 3C) to obtain two different measures of signal strength including: detection of peak voltage activity to cued eyeblinks and signal RMS energy to resting state activity. For eyeblink detection, subjects were guided via auditory cues to blink 10 times. Voltage peaks exceeding +100 μV within a 1-second window of the auditory cue were scored as a detected blink. A successful blink test was defined as 7 or more detect blinks. Following the blink test, the app interface showed a simplified graphical representation of the headband recording channels' 5-second broadband RMS signal strength. Channels with RMS signal strength ≤1 μV were shown in red, signal strength >1 μV and ≤5 μV in orange, and >5 μV in green.

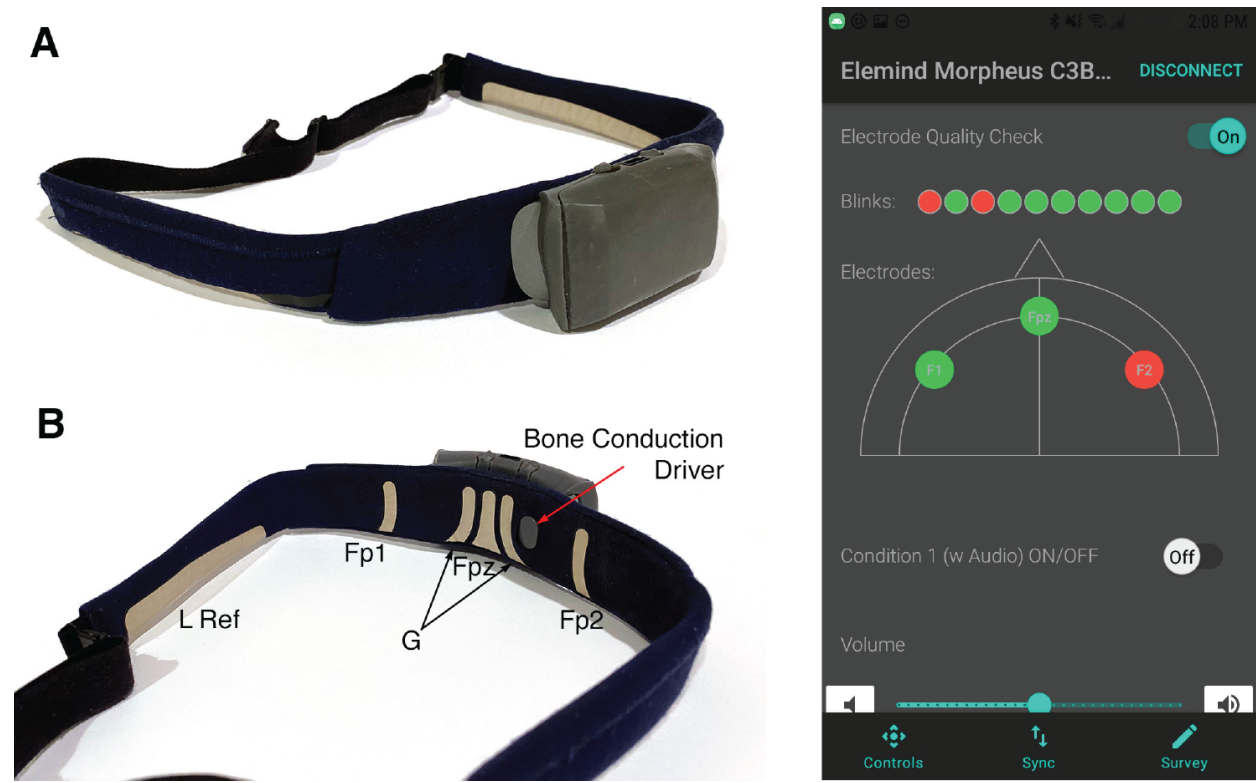

Figure 3: The Elemind Neuromodulation (ENMod) device. A) Front ¾ view of the processor unit ("puck") attached to the Muse S (2nd gen) headband. Left over-the-ear reference electrode shown. B) Back view showing the left over-the-ear reference electrode (L Ref), recording electrodes Fp1, Fpz, and Fp2, ground electrode (G), and the bone conduction driver. C) The screenshot of the smartphone application interface after a successful (8 out of 10 identified blinks) and 2 out of 3 scalp electrodes (Fp1 and Fpz) with adequate signal strength (>5 μV RMS).

Data were uploaded and analyzed once all study equipment was returned back to Elemind at the end of the study. Although the headband did not have the ability to upload the large EEG data logs remotely, it could remotely upload small (1 KB) metadata files to a secure cloud server detailing data log recording times and file sizes. Research staff monitored study progress through these metadata logs. Subjects were contacted directly when any lapse in protocol or schedule was suspected.

*Real-time phase estimation with the endpoint-corrected Hilbert Transform (ecHT)*: Our custom endpoint-corrected Hilbert Transform (ecHT) algorithm was implemented on the tabletop and wearable devices. This computation estimates the instantaneous phase of an oscillation at the leading edge of a sample window, overcoming the phase error inherent to existing approaches, such as the Hilbert Transform or the phase-locked loop . The methods and error corrections made possible with the ecHT computation have been described in detail previously (Schreglmann et al., 2021). Briefly, the EEG signal is sampled and decomposed into spectral components with a fast fourier transform (FFT). A causal bandpass filter is used to filter the frequency-domain representation of the data to correct for endpoint effects. Then an inverse FFT is used to compute the complex-valued analytic signal. At this step, the instantaneous phase and amplitude of the most recent sample is equal to the phase and magnitude of the last complex number in the analytic signal (Figure 4).

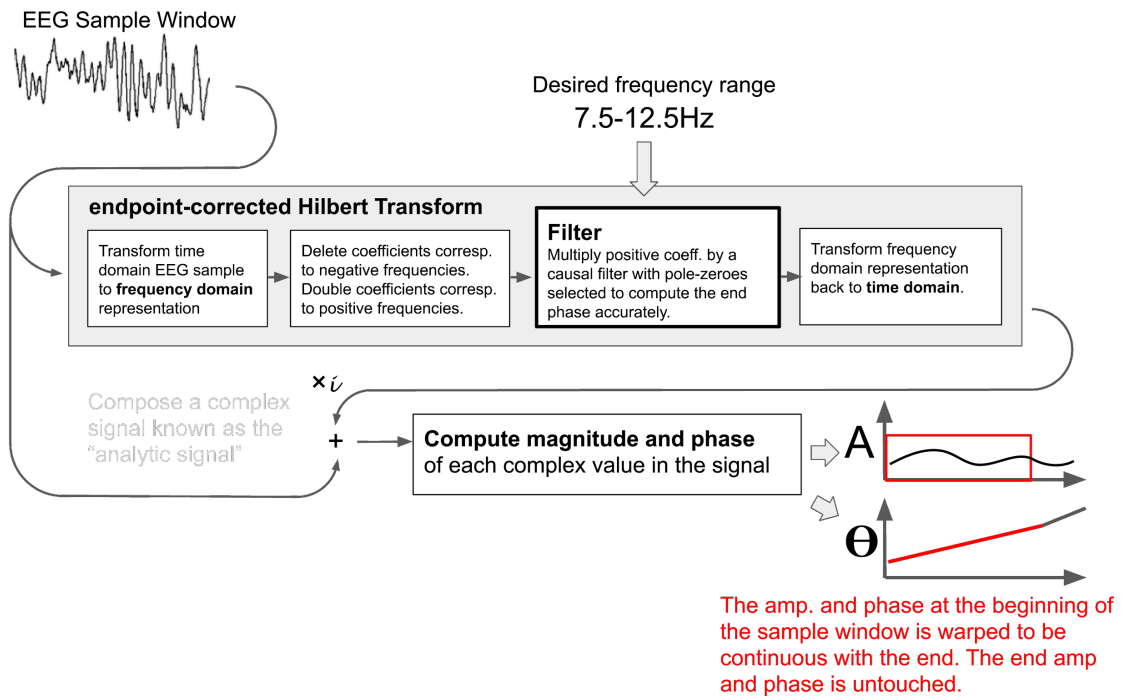

Figure 4: THE ENDPOINT-CORRECTED HILBERT TRANSFORM (ecHT).  This figure summarizes how HT is modified to generate the ecHT.  The ecHT uses an infinite impulse response causal bandpass filter, whose parameters (gain, poles, and zeros) were originally empirically optimized on sine-waves to preserve the phase of the most recent sample point.

*Auditory stimulation for evoked response measures:* To estimate population mean ERP latencies and effects on alpha oscillations, EEG was recorded in a sound isolated room and sound delivery was controlled by the tabletop ENMod device with audio output through Etymotic ER-4P earphones (www.etymotic.com). Short duration (12 ms), high intensity (85 dB) pink noise sound pulses were played at random phases relative to alpha with variable inter-stimulus intervals. To assess the phase-dependent effects on alpha oscillations, short duration (12 ms), high intensity (85 dB), pink noise sound pulses were played at peak and trough phases aligning the early ERP (P1) component to the individual subject's alpha trough and peak phase, respectively, with variable inter-stimulus intervals. Details of this computation are described in the Supplementary Figures and Data.

*Auditory stimulation for in-home sleep study.* Auditory stimulation on the wearable device used in at-home testing was provided through a bone conduction driver with 22 kHz mono WAV playback positioned in the middle of the user's forehead. The phase-locked auditory stimulus the subjects received was a combination of phase-locked pink noise sound pulses combined with a background natural rain sound. Through prior pilot studies, we determined that subjects best tolerated the high intensity pink noise stimulation when it was presented against a background of a gentle rain sound (Light of Mind ©). The phase-locked pink noise sounds were played at 18 dB above the background sound to ensure minimal masking and an effective signal-to-noise ratio. Accordingly, the sound pressure level (SPL) calculated from the root mean square (RMS) energy was 47 dB and 65 dB for the background rain and pink noise pulses, respectively. This sound combination maintains alpha phase-locking precision (Fig. 7). Because the perceived loudness of the audio depended on both the tightness of the headband fit and background noise levels, subjects were instructed to adjust the volume of the background rain until it was just audible. The pink noise phase-locking was determined by the instantaneous phase of the subject's alpha oscillations, such that the onset and offset of the stimulus appeared at 134° to 224° (alpha trough) and 314° to 44° (alpha peak). The two different phase conditions were based on timing of the P1 peak component of the evoked response to a single 12-ms pink noise burst such that the auditory evoked response arrived in-phase or out-of-phase with a 10-Hz alpha oscillation (Figure 5, also see Supplementary Data).

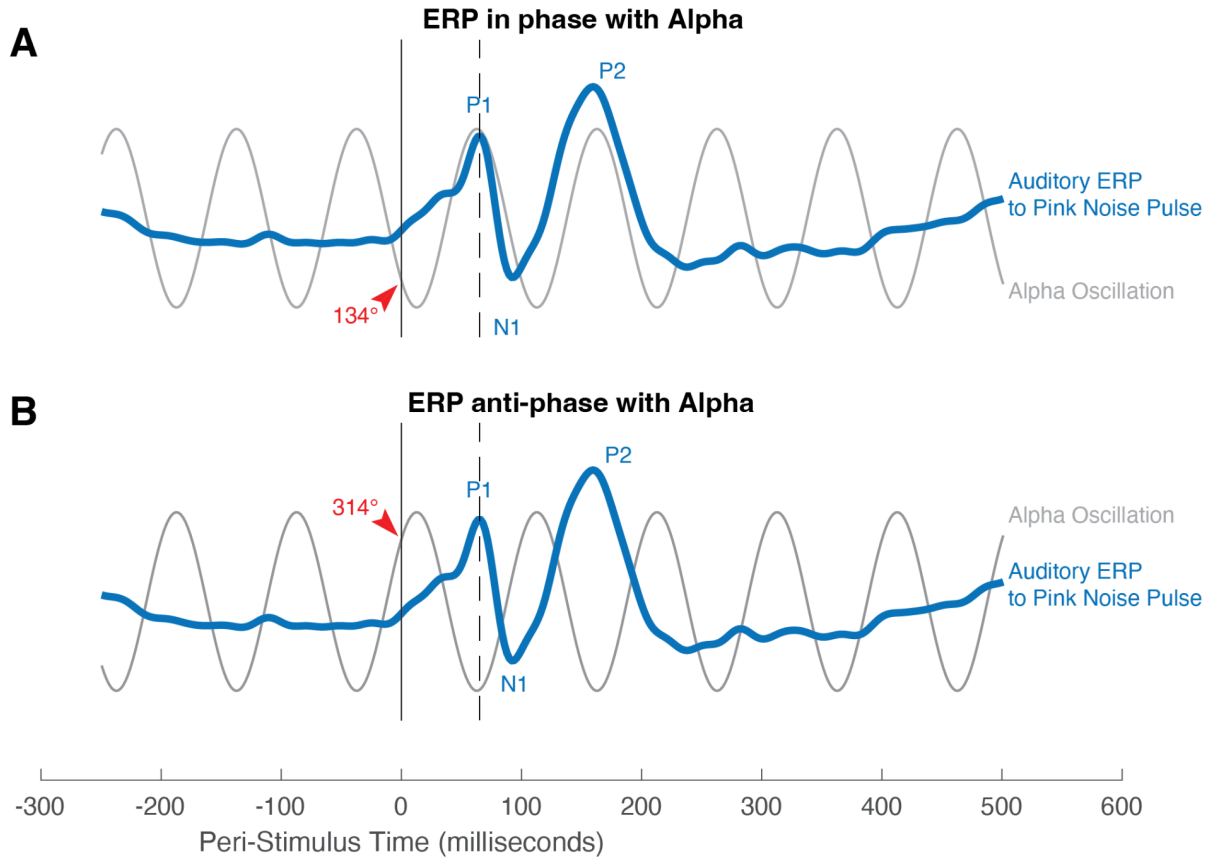

Figure 5: Early ERP arrival in-phase or anti-phase with the alpha oscillation. This schematic illustrates how the auditory stimulus is delivered to align the early (P1) auditory evoked response potential (ERP) in-phase or anti-phase with the subsequent peak or trough of alpha, respectively. . Aligning the ERP latency with these two target phases of alpha required us to estimate the ERP latency as well as the alpha frequency (IA . A) Accounting for an average ERP-P1 latency of 62 ms, the auditory stimulus was delivered at a phase of 134° so the ERP-P1 component coincided with the alpha peak. B) Accounting for the same ERP-P1 latency of 62 ms, an auditory stimulation at 314° produced an ERP-P1 component coincided with alpha trough phase.

*Sleep Study Data analysis:* Sleep staging was performed by a 3rd-party registered polysomnographic technologist (Sleep Strategies, Inc. Ottawa, Canada). EEG data logs were qualitatively assessed and ranked for signal quality. These data logs were subjectively sorted by quality, data conditions blinded to the scorer and delivered in random order in batches of 10 data logs. Visual scoring was complete/stopped when both the registered sleep tech and Elemind staff scientist agreed that the signal quality of the EEG data no longer contained information consistent with the AASM scoring guidelines. For our analysis, weekly average sleep onset latency times to the first epoch of N2 sleep were calculated from all available datasets (i.e., not all subjects had a complete complement of 4 nights of EEG data per condition).

**Results**

*EEG signals collected from the ENMod headband showed measurable neural oscillations and sleep microevents that were suitable for tracking sleep stage.* We first assessed the ability of our ambulatory EEG system to record neurological signals with sufficient fidelity for effective measurement of sleep-related activity. Datasets collected from sleeping subjects showed typical spectrotemporal features of sleep, including oscillatory activity in the alpha, beta, theta, and delta bands. Additionally, sleep-related microevents were also visible. We observed activity characteristic of sleep spindles, k-complexes and vertex waves (Figure 6). Though spectral distortions were observed at approximately 2 Hz due to a high-pass filter, data were suitable for standard visual sleep stage scoring by independent certified technicians.

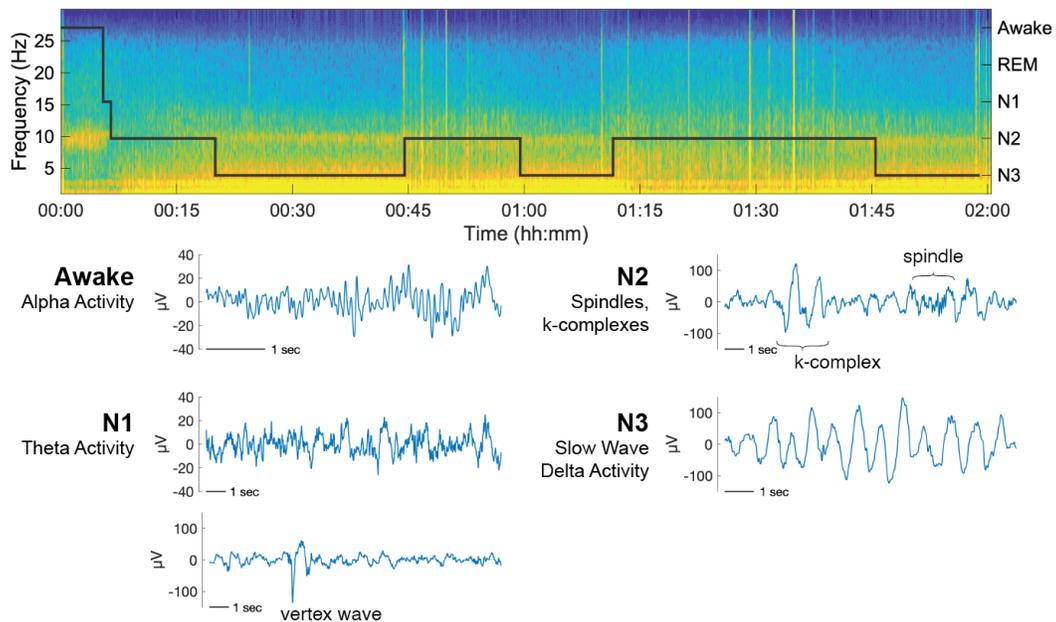

Figure 6: The wearable device records EEG with sufficient fidelity to distinguish oscillatory activity characteristic of sleep. Top) Time-frequency plot of spectral power (yellow max, blue minimum) recorded in the first two hours of overnight sleep with the wearable headband in one subject.  The overlaid hypnogram ( black line, y-axis on right) indicates four  sleep stages scored by a clinically certified sleep technician.  Sleep stages include Awake (eyes closed) behavior, Rapid Eye Movement (REM), and three progressively deeper non-REM stages of sleep (i.e., stages 1, 2, 3 or N1, N2, N3). Bottom) Time-voltage signals were suitable for standard visual sleep scoring, as shown for data from the same recording shown above.   "Awake" stage was dominated by EEG oscillations in the alpha (8-10 Hz) spectral range.  Stage-1 (N1) sleep was associated with increased theta (4-7 Hz) activity (top)  and vertex waves (bottom).  Stage-2 (N2) sleep was associated with increased incidence of k-complex and spindle oscillations micro-events. Stage-3 deep sleep (N3) was dominated by delta slow wave (0.5-4 Hz) oscillations.

Some datasets collected by subjects at home did not yield high-quality data. These datasets were typically characterized by broadband spectral activity measured on all 3 electrodes or a lack of any obvious neural activity. Of the 257 recorded datasets, 77 were found to be unusable due to signal quality issues (Table 2). Hardware malfunctions were ruled out in the majority of cases; therefore, these issues likely resulted from poor electrode contact with the subjects' scalp or degradation of the electrode fabric material. We observed similar kinds of data loss on specific channels on our comparator commercial system when the associated electrodes became disconnected while the subject was asleep, suggesting that loss of sufficient electrode contact was at issue in both cases.

To test the full-night performance of the system, we collected data from two subjects who wore the ENMod device with EEG recording active from bedtime until the subjects woke up (8 hours, 4 minutes of recording for one subject, and 8 hours, 15 minutes for the other). Sleep recordings were broken down into standard 30-second epochs and scored by a third-party Registered Polysomnographic Technologist (RPSGT). Of 1,958 epochs, 1,858 (95%) were deemed scorable, with the remaining 100 epochs (5%) receiving a classification of "Indeterminate." The distribution of sleep stages for one full overnight session is shown in Figure S2. Overall, these data demonstrate that our EEG system was able to capture sleep-related neural activity with sufficient fidelity for manual sleep scoring by an RPSGT.

*Table 2: Tabulation of datasets from all nights (288 data collection nights) of sleep recorded during the at-home study*

| Description | N | % | Description | N | % |
|---|---|---|---|---|---|
| Recorded data logs (out of 288) | 257 | 89.2% | Scored N1 onsets | 158 | 87.8% |
| Missing data logs [1] | 31 | 10.8% | Scored N2 onsets | 177 | 98.3% |
| RPSGT scored [2] | 180 | 70.0% | Scored N3 onsets | 178 | 98.9% |
| Unscorable [3] | 77 | 30.0% | Scored REM onsets | 10 | 5.6% |

[1] 31 of 288 data logs were never recorded due to user error or technical failures
[2] 180 of 257 data logs were of sufficient quality for a registered sleep technician to score
[3] 77 of 257 recorded data logs had insufficient data quality for sleep scoring

*On-device implementation of ecHT results in minimal phase error when tracking fast frequency neural oscillations*. To evaluate the performance of the ecHT algorithm for tracking phase with minimal latency, we computed the phase precision of our estimate using a post-hoc analysis of recorded EEG activity. Figure 7 A shows the phase accuracy of our phase-locked pink noise onset and offset events for a single representative 30-minute stimulation session targeting the peak phase of alpha. The average phase error for the onset stimulus events was -1.07° ± 43.8° and -8.87° ± 49.2° for offset events (mean ± SD). Figures 7 B and C summarize the across-session average phase accuracy for stimulation to alpha at trough and peak phase. For stimulation targeting the trough phase of alpha (Figure 7 B), the average across-session (n = 86 sessions) phase error was -5.39° ± 12.6° for onset events and -15.5° ± 18.7° for offset events (mean ± SD). For stimulation targeting the peak phase of alpha (Figure 7 C), the average across-session (n = 88 sessions) phase error was -4.04° ± 25.7° for onset events and -12.1° ± 29.7° for offset events (mean ± SD).

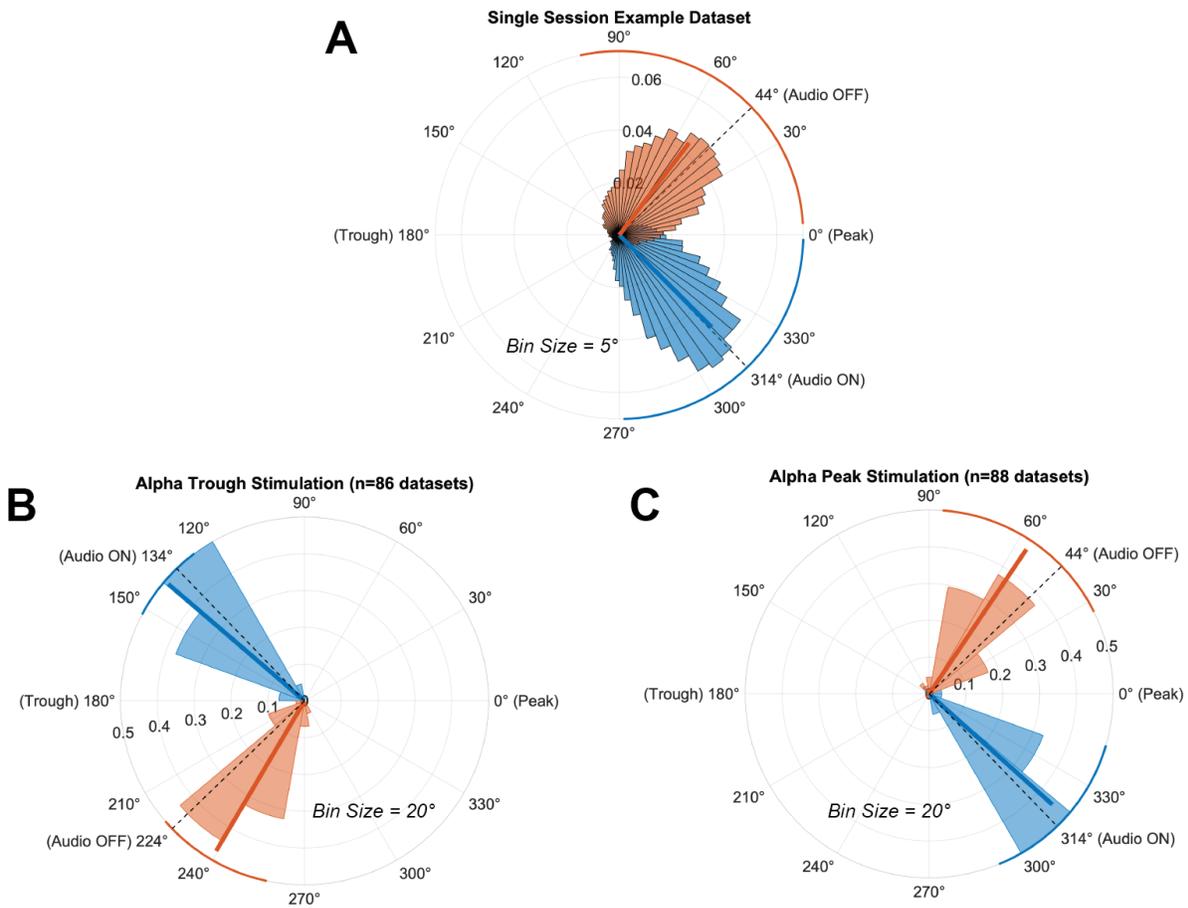

Figure 7: **A)** Distribution of target onset (blue: 314°) and offset (red: 44°) phases for pink noise pulses near the alpha peak phase for a single 30-minute phase-locked stimulation interval. Mean phase angles are represented by the solid blue (onset) and red (offset) vector lines; arcs at the edge of the polar axis represent one standard deviation. The phase locking value, a normalized metric of phase coherence, is represented by the length of the vector line relative to the radius of the polar axis. **B)** Session-averaged (n = 86 datasets) mean onset and offset phase values for stimulation targeting the trough phase of the alpha cycle. **C)** Session-averaged (n = 88 datasets) mean onset and offset phase values for stimulation targeting the peak phase of the alpha cycle. The radial axis labels denote the probability distribution values for the radial histogram bins. Bin size for Figure 7A = 5°, Figure 7B = 20°.

*Table 3a: Single Session Phase Accuracy Summary*

| Description | $N_{events}$ | Angle Mean ± SD | Median | PLV | Error Mean ± SD |
|---|---|---|---|---|---|
| Alpha Peak Onset (314°) | 16,374 | 315.1° ± 43.76° | 314.8° | 0.7084 | -1.073° ± 43.76° |
| Alpha Peak Offset (44°) | 16,374 | 52.87° ± 49.42° | 47.97° | 0.6281 | -8.865° ± 49.42° |

*Table 3b: Across-Session Phase Accuracy Summary*

| Description | $N_{datasets}$ | Angle Mean ± SD | Median | PLV | Error Mean ± SD |
|---|---|---|---|---|---|
| Alpha Trough Onset (134°) | 88 | 139.4° ± 12.59° | 139.1° | 0.9759 | -5.391° ± 12.59° |
| Alpha Trough Offset (224°) | 88 | 239.5° ± 18.65° | 239.1° | 0.9471 | -15.51° ± 18.65° |
| Alpha Peak Onset (314°) | 86 | 318.0° ± 25.72° | 314.7° | 0.8992 | -4.035° ± 25.72° |
| Alpha Peak Offset (44°) | 86 | 56.11° ± 29.67° | 56.14° | 0.9471 | -12.11° ± 29.67° |

***Auditory stimulation locked to alpha oscillations alters alpha activity in a phase-dependent manner***. Based on known neurophysiology, we hypothesized that audio stimulation would differentially modulate alpha when the early ERP component (P1) arrives during the excitable trough phase compared to the inhibited peak phase of alpha (Canolty & Knight, 2010; Fries et al., 2001; Haegens et al., 2015; Figure 5). In an initial session, the latency of the ERP P1 was measured in response to pink-noise sound pulses delivered at random phases with eyes open (Methods, Figure 5). In a second session, the alpha spectral band evoked response oscillation (ERO) was measured by using the predetermined ERP latency to deliver alpha peak or trough phase-locked pink-noise sound pulses. For both sessions, the auditory stimulus ERP P1 component latency was the same 64.2 ± 8.8 ms (Methods). On average, alpha oscillation amplitude following auditory stimulation with the peak phase-locked sound (Figure 8B, black line) was reduced relative to the alpha oscillation amplitude with trough phase-locked sound stimulation (Figure 8B, green line). These data provided a population estimate of average pink-noise sound evoked potential latencies to use in the sleep study in order to influence alpha. Moreover, this supported the hypothesis that auditory stimulation modulates alpha oscillations in a phase-dependent manner.

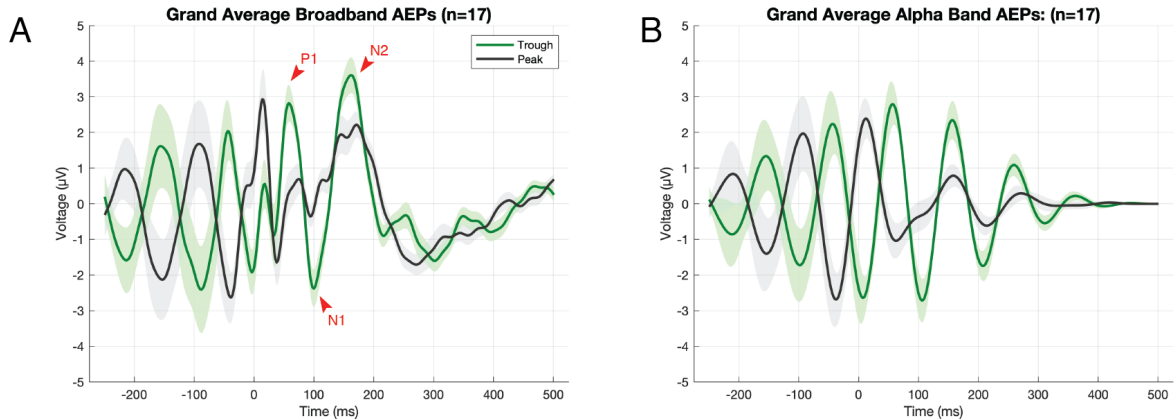

Figure 8: Stimulation timing relative to phase of alpha rhythms alters subsequent alpha wave amplitude. A) The P50 and N100 components of the grand average broadband auditory evoked potentials (AEPs) are reduced when pink noise pulses are presented at the trough phase of the alpha cycle (black). B) Bandpassed AEPs to the alpha band (7.5-12 Hz) show stimulation to alpha peak disrupts the post-stimulus alpha oscillations. Stimulation at the trough of the pre-stimulus alpha cycle does not impede the subsequent post-stimulus alpha cycles.

*Closed-loop auditory stimulation effects on sleep onset latency.* Alpha oscillations are elevated in wakefulness and throughout multiple sleep stages in subjects with insomnia disorders including those with sleep initiation problems *(Z*(Riedner et al., 2016; Zhao et al., 2021)*).* Given the observation that peak and trough phase-locked sounds differentially impacted alpha oscillations (Figure 8), this feasibility study set out to compare stage-2 sleep onset latency (SOL-N2) under three sleep conditions including: 1) no sound stimulation; 2) peak (314°) and 3) trough (134°) phase-locked sound stimulation. To minimize potential effects of home environment sounds and to make the sounds more soothing, phase-locked pink-noise sounds were played jointly with background natural rain sound at a high (18 dB) signal-to-noise ratio. Subjects were screened and included in the study based on scores assessed with the insomnia severity index, Pittsburgh sleep quality index, and self-report that it took them 30 minutes or more to fall asleep regularly (Table 1). Following screening, 24 subjects (11 male, 13 female; Age: 33.0 ± 6.6 yrs (mean ± SD), Median age: 31 yrs, Range [26 55] yrs) completed a 3-week, randomized, cross-over in-home sleep study (Figure 9). In a video meeting, study personnel walked subjects through the daily routine for operating the wearable headband device with a smartphone application (Figure 3). On randomly assigned weeks with either type of experimental phase-locked auditory stimulation, the sounds were played for 30 minutes during sleep initiation.

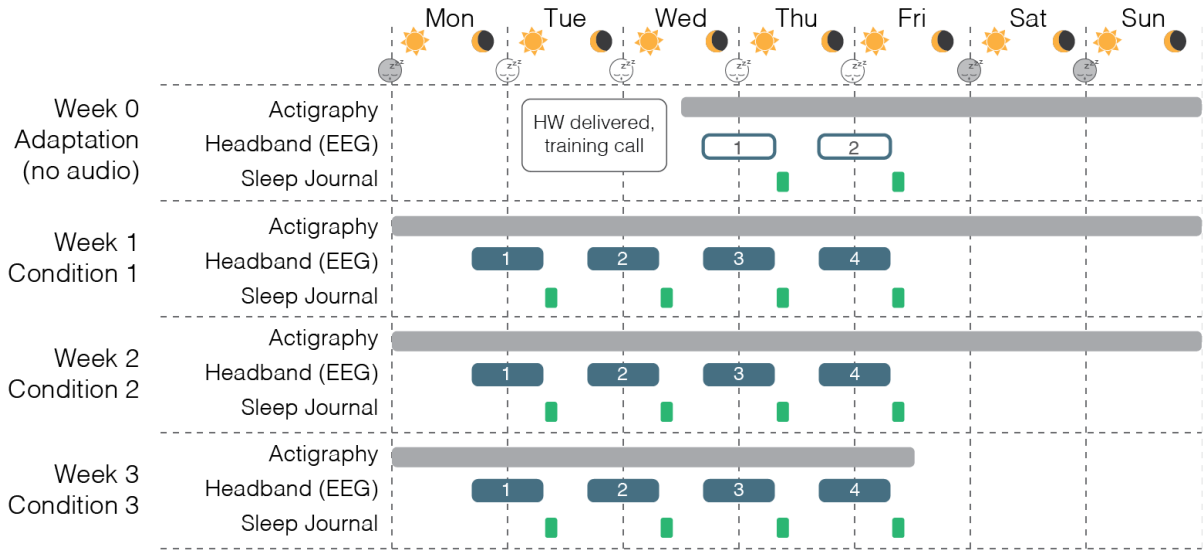

Figure 9: At-home sleep study design. Subjects wore the headband without audio stimulation or EEG data recording for two Adaptation nights prior to data collection. Four nights (Mon-Thu) of EEG data were collected each week for 3 weeks of testing across 3 separate conditions: Control (no audio stimulation), phase-locked stimulation to alpha peak, and phase-locked stimulation to alpha trough. Conditions were randomized and counter-balanced across weeks. Actigraphy data collection began on the day of the first adaptation night and continued throughout the duration of the study.

To quantify the sleep stages and sleep onset latency outcome measures, de-identified EEG data was scored by independent sleep scoring technicians across all recording nights. For data pooled across all (n=24) subjects, there was no significant difference in stage-2 sleep onset latency (SOL-N2) across the three experimental conditions (no audio: 16.0 ±14.6 mins; trough phase (134°) 13.3 ±8.1 mins; peak phase (314°) : 12.3 ±8.7 mins; one-way ANOVA(2,66) = 0.9938, p = 0.3756; Figure 10 A). Despite our inclusion criteria, 71% (n=17 of 24) of subjects did not display objective EEG verified sleep onset latencies greater than 30 minutes. The lack of effect on the full population of participants could reflect the variation in sleep initiation problems across weeks or alternatively a subjective misperception of sleep onset problems in 71% of the subjects included here (Rezaie et al., 2018). In contrast, 29% (n=7) of subjects were objectively defined as "poor sleepers" with sleep initiation problems confirmed by at least one night of EEG verified SOL-N2 of 30 minutes or more. During weeks when the ENMod headband was programmed to deliver alpha phase-locked sound stimulation, these "poor sleepers" had a significantly reduced average SOL-N2. A one-way analysis of variance found weekly average SOL-N2 f was significantly reduced by phase-locked auditory stimulation versus no-audio conditions (F(2,18) = 5.8478, p = 0.011; Figure 10 B). Tukey's HSD test for multiple comparisons found that phase-locked stimulation to both the peak cycle of alpha (p = 0.0125, 95% C.I. = [-35.2614, -4.0958]) and trough cycle of alpha (p = 0.0453, 95% C.I. = [-31.4697 -0.3041]) were significantly different from the control condition without stimulation. There was no statistically significant difference between the peak and trough stimulation conditions (p = 0.8106). Without stimulation, average weekly sleep onset time was 35.3 (± 13.0) minutes. With phase-locked auditory stimulation to alpha trough and peak sleep onset time was 19.4 (± 10.8) and 15.6 (±10.3) minutes, reflecting a reduction in SOL-N2 by 15.9 (±22) and 19.7 (± 20.4) minutes, respectively (Figure 10 B).

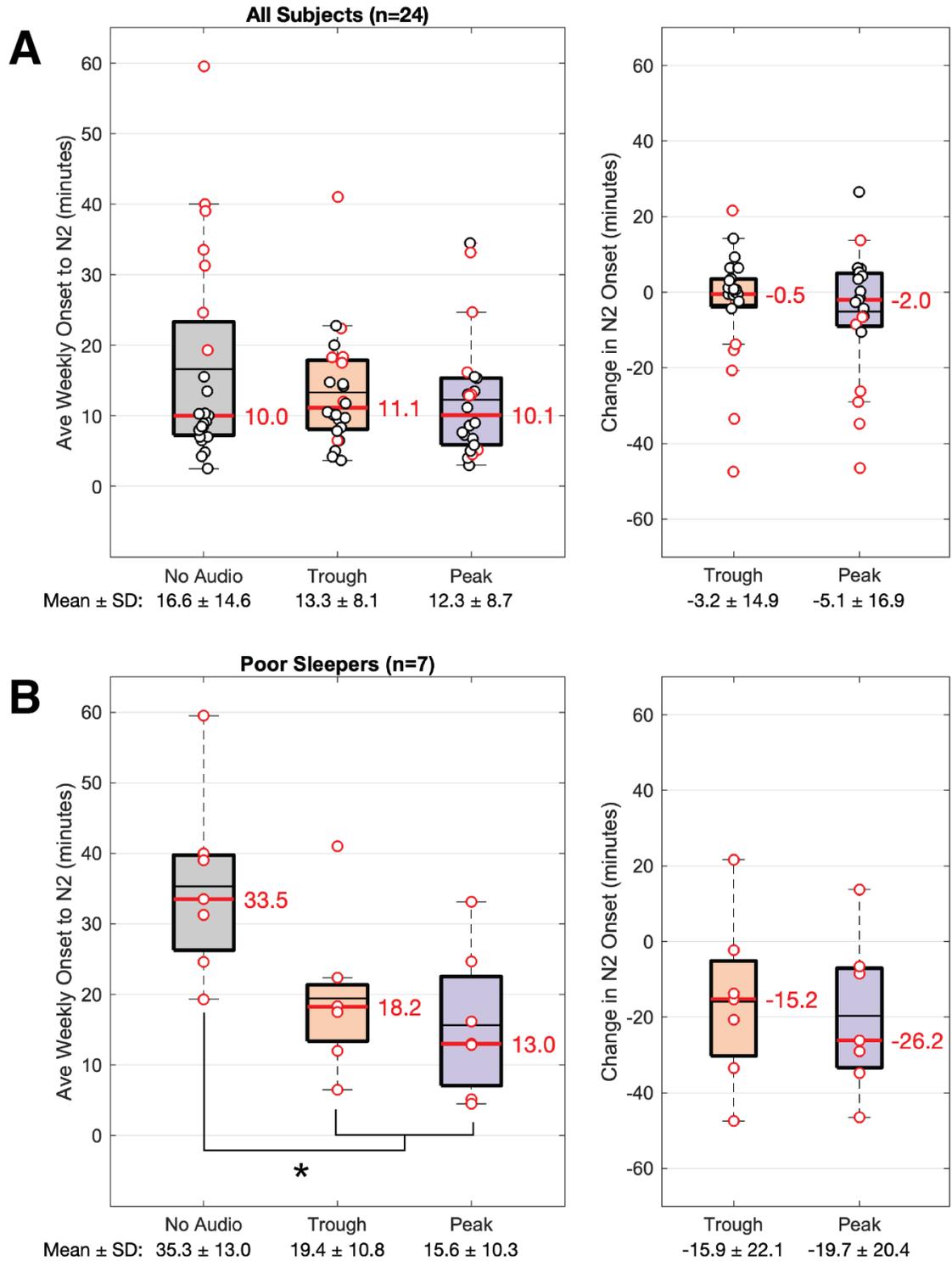

Figure 10: Changes in sleep onset latency for subjects in Control, Trough, or Peak stimulation conditions. A) Sleep onset latencies for all subjects with scorable datasets (n=24). Box boundaries show the interquartile range, while red lines depict the median. Black lines represent the means of each population. B) Reduction in sleep onset latencies for each stimulation condition relative to control. Markings are as in A. C and D): Sleep onset latencies for individuals with at least one night in which sleep onset latency was measured to be greater than 30 mins by visual EEG scoring (n=7). * p < 0.05

While onset times to non-REM Stage 2 (N2) sleep were improved with phase-locked auditory stimulation, the other two measures of sleep onset latency (actigraphy and subjective morning surveys) failed to reflect similar improvements. Weekly average sleep onset as measured by actigraphy was 19.2 ± 13.1 minutes for the baseline (no audio) condition, 13.2 ± 6.5 minutes for phase-locked stimulation to alpha trough (134°), and 10.2 ± 7.6 minutes for stimulation to alpha peak (314°; Figure S4). A one-way analysis of variance confirmed no significant effect of stimulation condition [F(2,18) = 1.6070, p = 0.2280]. Weekly average subjectively-reported sleep onset times were 30.2 ± 10.4 minutes the baseline (no audio) condition, 25.3 ± 7.5 minutes for stimulation to alpha trough, and 21.4 ± 8.9 minutes for stimulation to alpha peak [one-way ANOVA: F(2,18) = 1.6598, p = 0.2180] (Figure S5).

**Discussion**

This feasibility study developed and tested a wearable device for accurate closed-loop auditory stimulation phase-locked to alpha oscillations. The device employed a novel endpoint-corrected Hilbert Transform algorithm implemented with efficient EEG signal processing and device control operations. First, these results demonstrate feasibility for real-time EEG signal processing technology to accurately track instantaneous alpha phase and deliver phase-locked audible sound stimulation to modulate alpha in a phase-dependent manner. This modulation is consistent with the neurophysiology of synchronous rhythmic neural oscillations such as alpha (Canolty & Knight, 2010; Fries et al., 2001; Haegens et al., 2015) and supports our hypothesis that sounds phase-locked to the peak of alpha such that the ERP arrives during the trough are more likely to disrupt alpha oscillations (Figure 8). Secondly, the current results demonstrate feasibility to implement this alpha phase-locking technology on a wearable, non-invasive, EEG-based neuromodulation device for home use. During the overnight sleep study, objective EEG-based measures of sleep onset latency (SOL-N2) were reduced for poor sleepers by up to ~20 minutes with peak phase-locked auditory stimulation versus a no auditory control condition (Figure 10B). Importantly, the magnitude of these effects are within the range of sleep onset latency reductions (aka, 10 to 17 minutes) observed across multiple studies with sleep hypnotics (e.g. zolpidem) versus placebo drugs (Xiang et al., 2021) and for an FDA certified thermal cooling headband versus placebo (Roth et al., 2018). This is our rationale for continued testing and refinement of phase-dependent audible sound stimulation for disrupting alpha and promoting healthy sleep initiation. Finally, subjective reports and wrist-worn accelerometry measures of sleep onset latency where not different across control "no audio" versus either phase-locked audio condition. However, subjective measures (Rezaie et al., 2018) and non-EEG-enabled wearables (Miller et al., 2022; Stone et al., 2020) have been shown to be poor estimates of sleep stage, highlighting the need for wearable devices capable of recording neural activity.

Beyond demonstrating the feasibility of this approach, this study yielded additional insight into features that could improve future iterations of wearable EEG devices to help ameliorate hyperarousal and promote healthy sleep. Although many of our subjects' recordings contained EEG signals that were scorable by an RPSGT, in our at-home study, we observed that 30% of datasets did not contain usable data. This outcome was primarily related to issues around poor contact between electrodes and the skin or degradation of the electrodes themselves. Without the ability to effectively record neural data, closed-loop approaches like the one presented here are not possible. Long-term durability or easy replacement of electrodes is therefore an important area of improvement, and several groups are currently working on this problem(Tseghai et al., 2021). Finally, 10.8% of subject-nights in our study did not yield datasets at all. Though it was hard to identify a single issue that resulted in these losses, many lost nights of data were related to user error. Designing neuromodulation systems that are intended for consumer use will need to address issues of usability and stability in an at-home setting. One such approach may be to provide easy-to-understand feedback about device function and performance, such as signal quality or electrode impedance, so that users can better understand how to achieve optimal results.

As part of the evaluation of our device, we performed an in-home sleep study in which subjects applied the device and recorded data on their own after a virtual training session with researchers. This study yielded insight into the usability of our system, as well as tested the use of closed-loop auditory neuromodulation as an intervention for accelerating sleep initiation. After restricting our analysis to datasets with scorable EEG data, we did not observe a significant effect of stimulation (either optimal or pessimal) on SOL across the whole subject population, relative to sham stimulation (no sound). Interestingly, despite self-report of symptomatic insomnia, including an Insomnia Severity Index (ISI) score of >21 and PSQI score > 5, the majority of our subjects did not display sleep initiation difficulties as scored by EEG. These observations are not inconsistent with what is known about insomnia patients, of which up to 50% display discrepancies between subjective reports of sleep and what can be measured objectively (Rezaie et al., 2018). Indeed, a recent study found that insomnia can be subdivided into at least 5 different subtypes with unique characteristics that remain stable over time (Blanken et al., 2019). In our data, we also observed that a subset of our subjects, specifically those with observable SOLs > 30 min, appeared to respond favorably to the stimulation protocol, while those with shorter SOLs did not. These differences in response may be reflective of unique features or etiologies underpinning the various subtypes of insomnia, and suggests the possibility that alpha rhythms may be more important in certain forms of insomnia than others. Alternatively, the lack of effect observed in subjects with low SOLs may be due to a floor effect, in which our approach cannot accelerate sleep onset below a certain threshold.

Our study included 3 test conditions: a control condition in which no audio was played, one condition in which auditory pulses were delivered at the optimal phase of alpha, and one where pulses were delivered at the pessimal phase of alpha. Based on previous observations that elevated alpha power is associated with delays in sleep onset, we expected that auditory stimulation delivered at the pessimal alpha phase would be the most efficacious for reducing sleep onset. However, in the subset of individuals that exhibited a reduction in SOL, both stimulation conditions appeared to be equivalent. We hypothesize three possible explanations for this outcome. The first is that the audio alone was sufficient to affect sleep onset in these subjects. While broadband sounds have been reported to decrease sleep onset latency in insomniacs, the magnitude of the effect is less than that reported here (Messineo et al., 2017).

Another possibility is that auditory stimulation, regardless of phase, disrupts alpha enough to accelerate sleep onset. Indeed, prior studies have shown that salient and high intensity sensory stimuli can decrease the late ERP components and desynchronize alpha oscillations independent of alpha phase (Fodor et al., 2020; Iemi et al., 2019; Mazaheri & Jensen, 2008; Nierhaus et al., 2009; Nikulin et al., 2007). Lastly, subtypes of insomniacs have been observed to differ in response to auditory-evoked neural responses (Blanken et al., 2019). Individuals may respond differently to stimulation at particular phases of alpha, and an approach tailored to individual phenotypes may be required to achieve the maximum effect. A better understanding of the relationship between elevated alpha power and sleep onset as well as additional experiments would be necessary to evaluate these hypotheses.

## Conclusion

While previous reports have identified closed-loop auditory stimulation as a potential mechanism for improving slow-wave sleep, ours is to our knowledge the first wearable device designed to provide closed-loop auditory stimulation timed to particular phases of fast (>8 Hz) oscillations with high phase precision. Our system was operable by subjects in an at-home setting, allowing individuals to administer therapy and collect data independently. Further developments will improve usability and possibly include the flexibility to tailor the stimulation to individual phenotypes. This approach may also have applications outside of sleep in conjunction with other stimulus modalities by targeting conditions associated with faster oscillations such as essential tremor (Schreglmann et al., 2021), or other conditions associated with measurable differences in neural oscillations (Takeuchi & Berényi, 2020). By enabling real-time analysis of oscillatory signals, wearable-based approaches such as this one could be explored as potential alternatives to invasive neuromodulation strategies.

## Author Contributions

Study concept and design: S.B., H.R., D.W. Fabrication and programming: D.W. Data acquisition: S.B., R.Y. Data analysis: S.B. Writing the manuscript: R.N., S.B. Revising the manuscript: S.B., R.N., H.R., R.Y.

## Ethical considerations

This research was conducted in accordance with the principles of the declaration of Helsinki as well as local regulations. Studies conducted by Elemind Technologies, Inc. were approved by an independent institutional review board (Solutions IRB, Yarnell, AZ). Studies conducted at the University of Connecticut were approved by the Institutional Review Board at the University of Connecticut. Participants were provided verbal and written descriptions of the experiment procedures and study rationale and they consented to participate in the experiments.

## Data availability statement

The data that support the findings of this study are available upon reasonable request from the authors.

**Competing interest statement**

Authors S.B., R.N., R.Y, and D.W. are employees of Elemind Technologies, Inc. H.L.R. has ownership interest in Elemind Technologies, Inc. Elemind did not sponsor the basic research included here from University of Connecticut.

**Funding**

This work was funded in part by Elemind Technologies, Inc. The work carried out by the Brain Computer Interface Core at University of Connecticut was supported by an Academic Plan grant from the University of Connecticut.

**Acknowledgements**

The authors would like to thank Drs. Nir Grossman, Ed Boyden, Ruth Benca and Derk-Jan Dijk for input into experimental design and fruitful discussions.

**References**

Berro, L. F., Overton, J. S., Reeves-Darby, J. A., & Rowlett, J. K. (2021). Alprazolam-induced EEG spectral power changes in rhesus monkeys: A translational model for the evaluation of the behavioral effects of benzodiazepines. *Psychopharmacology*, *238*(5), 1373–1386. https://doi.org/10.1007/s00213-021-05793-z

Blanken, T. F., Benjamins, J. S., Borsboom, D., Vermunt, J. K., Paquola, C., Ramautar, J., Dekker, K., Stoffers, D., Wassing, R., Wei, Y., & Van Someren, E. J. W. (2019). Insomnia disorder subtypes derived from life history and traits of affect and personality. *The Lancet Psychiatry*, *6*(2), 151–163. https://doi.org/10.1016/S2215-0366(18)30464-4

Buscemi, N., Vandermeer, B., Friesen, C., Bialy, L., Tubman, M., Ospina, M., Klassen, T. P., & Witmans, M. (2007). The Efficacy and Safety of Drug Treatments for Chronic Insomnia in Adults: A Meta-analysis of RCTs. *Journal of General Internal Medicine*, *22*(9), 1335–1350. https://doi.org/10.1007/s11606-007-0251-z

Canolty, R. T., & Knight, R. T. (2010). The functional role of cross-frequency coupling. *Trends in Cognitive Sciences*, *14*(11), 506–515. https://doi.org/10.1016/j.tics.2010.09.001

Carney, C. E., Buysse, D. J., Ancoli-Israel, S., Edinger, J. D., Krystal, A. D., Lichstein, K. L., & Morin, C. M. (2012). The consensus sleep diary: Standardizing prospective sleep self-monitoring. *Sleep*, *35*(2), 287–302. https://doi.org/10.5665/sleep.1642

Debellemaniere, E., Chambon, S., Pinaud, C., Thorey, V., Dehaene, D., Léger, D., Chennaoui, M., Arnal,

P. J., & Galtier, M. N. (2018). Performance of an Ambulatory Dry-EEG Device for Auditory Closed-Loop Stimulation of Sleep Slow Oscillations in the Home Environment. *Frontiers in Human Neuroscience*, *12*. https://www.frontiersin.org/articles/10.3389/fnhum.2018.00088

Edinger, J. D., Arnedt, J. T., Bertisch, S. M., Carney, C. E., Harrington, J. J., Lichstein, K. L., Sateia, M. J., Troxel, W. M., Zhou, E. S., Kazmi, U., Heald, J. L., & Martin, J. L. (2021). Behavioral and psychological treatments for chronic insomnia disorder in adults: An American Academy of Sleep Medicine clinical practice guideline. *Journal of Clinical Sleep Medicine: JCSM: Official Publication of the American Academy of Sleep Medicine*, *17*(2), 255–262. https://doi.org/10.5664/jcsm.8986

Ferster, M. L., Da Poian, G., Menachery, K., Schreiner, S. J., Lustenberger, C., Maric, A., Huber, R., Baumann, C. R., & Karlen, W. (2022). Benchmarking Real-Time Algorithms for In-Phase Auditory Stimulation of Low Amplitude Slow Waves With Wearable EEG Devices During Sleep. *IEEE Transactions on Biomedical Engineering*, *69*(9), 2916–2925. https://doi.org/10.1109/TBME.2022.3157468

Fodor, Z., Marosi, C., Tombor, L., & Csukly, G. (2020). Salient distractors open the door of perception: Alpha desynchronization marks sensory gating in a working memory task. *Scientific Reports*, *10*(1), 19179. https://doi.org/10.1038/s41598-020-76190-3

Fries, P., Reynolds, J. H., Rorie, A. E., & Desimone, R. (2001). Modulation of oscillatory neuronal synchronization by selective visual attention. *Science (New York, N.Y.)*, *291*(5508), 1560–1563. https://doi.org/10.1126/science.1055465

Grossman, N., Wang, D., & Boyden, E. (2017). *Methods and Apparatus for Neuromodulation* (United States Patent No. US20170020447A1). https://patents.google.com/patent/US20170020447A1/en

Haegens, S., Barczak, A., Musacchia, G., Lipton, M. L., Mehta, A. D., Lakatos, P., & Schroeder, C. E. (2015). Laminar Profile and Physiology of the α Rhythm in Primary Visual, Auditory, and Somatosensory Regions of Neocortex. *The Journal of Neuroscience: The Official Journal of the Society for Neuroscience*, *35*(42), 14341–14352. https://doi.org/10.1523/JNEUROSCI.0600-15.2015

Hirshkowitz, M. (2016). Polysomnography Challenges. *Sleep Medicine Clinics*, *11*(4), 403–411. https://doi.org/10.1016/j.jsmc.2016.07.002

Huang, G., Liu, J., Li, L., Zhang, L., Zeng, Y., Ren, L., Ye, S., & Zhang, Z. (2019). A novel training-free externally-regulated neurofeedback (ER-NF) system using phase-guided visual stimulation for alpha modulation. *NeuroImage*, *189*, 688–699. https://doi.org/10.1016/j.neuroimage.2019.01.072

Iemi, L., Busch, N. A., Laudini, A., Haegens, S., Samaha, J., Villringer, A., & Nikulin, V. V. (2019). Multiple mechanisms link prestimulus neural oscillations to sensory responses. *ELife*, *8*, e43620. https://doi.org/10.7554/eLife.43620

Kang, S.-G., Mariani, S., Marvin, S. A., Ko, K.-P., Redline, S., & Winkelman, J. W. (2018). Sleep EEG spectral power is correlated with subjective-objective discrepancy of sleep onset latency in major depressive disorder. *Progress in Neuro-Psychopharmacology & Biological Psychiatry*, *85*, 122–127. https://doi.org/10.1016/j.pnpbp.2018.04.010

Klimesch, W. (1999). EEG alpha and theta oscillations reflect cognitive and memory performance: A review and analysis. *Brain Research Reviews*, *29*(2), 169–195. https://doi.org/10.1016/S0165-0173(98)00056-3

Lozano-Soldevilla, D. (2018). On the Physiological Modulation and Potential Mechanisms Underlying Parieto-Occipital Alpha Oscillations. *Frontiers in Computational Neuroscience*, *12*. https://www.frontiersin.org/articles/10.3389/fncom.2018.00023

Mazaheri, A., & Jensen, O. (2008). Asymmetric amplitude modulations of brain oscillations generate slow evoked responses. *The Journal of Neuroscience: The Official Journal of the Society for Neuroscience*, *28*(31), 7781–7787. https://doi.org/10.1523/JNEUROSCI.1631-08.2008

Messineo, L., Taranto-Montemurro, L., Sands, S. A., Oliveira Marques, M. D., Azabarzin, A., & Wellman, D. A. (2017). Broadband Sound Administration Improves Sleep Onset Latency in Healthy Subjects in a Model of Transient Insomnia. *Frontiers in Neurology*, *8*, 718. https://doi.org/10.3389/fneur.2017.00718

Miller, D. J., Sargent, C., & Roach, G. D. (2022). A Validation of Six Wearable Devices for Estimating Sleep, Heart Rate and Heart Rate Variability in Healthy Adults. *Sensors (Basel, Switzerland)*, *22*(16), 6317. https://doi.org/10.3390/s22166317

Morin, C. M., Belleville, G., Bélanger, L., & Ivers, H. (2011). The Insomnia Severity Index: Psychometric indicators to detect insomnia cases and evaluate treatment response. *Sleep*, *34*(5), 601–608.

Ngo, H.-V. V., Miedema, A., Faude, I., Martinetz, T., Mölle, M., & Born, J. (2015). Driving sleep slow

oscillations by auditory closed-loop stimulation-a self-limiting process. *The Journal of*

*Neuroscience: The Official Journal of the Society for Neuroscience*, *35*(17), 6630–6638.

https://doi.org/10.1523/JNEUROSCI.3133-14.2015

Nierhaus, T., Schön, T., Becker, R., Ritter, P., & Villringer, A. (2009). Background and evoked activity and their interaction in the human brain. *Magnetic Resonance Imaging*, *27*(8), 1140–1150.

https://doi.org/10.1016/j.mri.2009.04.001

Nikulin, V. V., Linkenkaer-Hansen, K., Nolte, G., Lemm, S., Müller, K. R., Ilmoniemi, R. J., & Curio, G.

(2007). A novel mechanism for evoked responses in the human brain. *The European Journal of*

*Neuroscience*, *25*(10), 3146–3154. https://doi.org/10.1111/j.1460-9568.2007.05553.x

Olfson, M., Wall, M., Liu, S.-M., Morin, C. M., & Blanco, C. (2018). Insomnia and Impaired Quality of Life

in the United States. *The Journal of Clinical Psychiatry*, *79*(5), 9151.

https://doi.org/10.4088/JCP.17m12020

Papalambros, N. A., Santostasi, G., Malkani, R. G., Braun, R., Weintraub, S., Paller, K. A., & Zee, P. C.

(2017). Acoustic Enhancement of Sleep Slow Oscillations and Concomitant Memory

Improvement in Older Adults. *Frontiers in Human Neuroscience*, *11*.

https://www.frontiersin.org/articles/10.3389/fnhum.2017.00109

Phillips, A. J. K., Vidafar, P., Burns, A. C., McGlashan, E. M., Anderson, C., Rajaratnam, S. M. W.,

Lockley, S. W., & Cain, S. W. (2019). High sensitivity and interindividual variability in the response

of the human circadian system to evening light. *Proceedings of the National Academy of*

*Sciences of the United States of America*, *116*(24), 12019–12024.

https://doi.org/10.1073/pnas.1901824116

Rezaei, M., Mohammadi, H., & Khazaie, H. (2019). Alpha-wave characteristics in psychophysiological

insomnia. *Journal of Medical Signals & Sensors*, *9*(4), 259.

https://doi.org/10.4103/jmss.JMSS_51_18

Rezaie, L., Fobian, A. D., McCall, W. V., & Khazaie, H. (2018). Paradoxical insomnia and

subjective–objective sleep discrepancy: A review. *Sleep Medicine Reviews*, *40*, 196–202.

https://doi.org/10.1016/j.smrv.2018.01.002

Riedner, B. A., Goldstein, M. R., Plante, D. T., Rumble, M. E., Ferrarelli, F., Tononi, G., & Benca, R. M. (2016). Regional Patterns of Elevated Alpha and High-Frequency Electroencephalographic Activity during Nonrapid Eye Movement Sleep in Chronic Insomnia: A Pilot Study. *Sleep*, *39*(4), 801–812. https://doi.org/10.5665/sleep.5632

Roth, T., Mayleben, D., Feldman, N., Lankford, A., Grant, T., & Nofzinger, E. (2018). A novel forehead temperature-regulating device for insomnia: A randomized clinical trial. *Sleep*, *41*(5), zsy045. https://doi.org/10.1093/sleep/zsy045

Schreglmann, S. R., Wang, D., Peach, R. L., Li, J., Zhang, X., Latorre, A., Rhodes, E., Panella, E., Cassara, A. M., Boyden, E. S., Barahona, M., Santaniello, S., Rothwell, J., Bhatia, K. P., & Grossman, N. (2021). Non-invasive suppression of essential tremor via phase-locked disruption of its temporal coherence. *Nature Communications*, *12*(1), Article 1. https://doi.org/10.1038/s41467-020-20581-7

Stone, J. D., Rentz, L. E., Forsey, J., Ramadan, J., Markwald, R. R., Jnr, V. S. F., Galster, S. M., Rezai, A., & Hagen, J. A. (2020). <p>Evaluations of Commercial Sleep Technologies for Objective Monitoring During Routine Sleeping Conditions</p>. *Nature and Science of Sleep*, *12*, 821–842. https://doi.org/10.2147/NSS.S270705

Takeuchi, Y., & Berényi, A. (2020). Oscillotherapeutics – Time-targeted interventions in epilepsy and beyond. *Neuroscience Research*, *152*, 87–107. https://doi.org/10.1016/j.neures.2020.01.002

Tseghai, G. B., Malengier, B., Fante, K. A., & Langenhove, L. V. (2021). The Status of Textile-Based Dry EEG Electrodes. *Autex Research Journal*, *21*(1), 63–70. https://doi.org/10.2478/aut-2019-0071

Vyazovskiy, V. V. (2015). Sleep, recovery, and metaregulation: Explaining the benefits of sleep. *Nature and Science of Sleep*, *7*, 171–184. https://doi.org/10.2147/NSS.S54036

Xiang, T., Cai, Y., Hong, Z., & Pan, J. (2021). Efficacy and safety of Zolpidem in the treatment of insomnia disorder for one month: A meta-analysis of a randomized controlled trial. *Sleep Medicine*, *87*, 250–256. https://doi.org/10.1016/j.sleep.2021.09.005

Zhao, W., Van Someren, E. J. W., Li, C., Chen, X., Gui, W., Tian, Y., Liu, Y., & Lei, X. (2021). EEG spectral analysis in insomnia disorder: A systematic review and meta-analysis. *Sleep Medicine Reviews*, *59*, 101457. https://doi.org/10.1016/j.smrv.2021.101457

**Supplementary Figures and Data**

**Computation of auditory stimulation timing for optimal and pessimal phase***: To determine the optimal and pessimal stimulation times, we measured the individual alpha center frequencies (IAF) and P50 latencies on a separate cohort of subjects (n=21) in a laboratory setting, using the benchtop version of our device. EEG was recorded from Fpz and Fz and sampled at 500 Hz. *Estimation of IAF*: Subjects were instructed to remain seated with eyes closed while EEG was passively recorded for two minutes. From the timeseries EEG data, we computed a 4-taper spectrogram using a 6-second sliding analysis window and step size of 150 ms (Figure S1A). From these data, the IAF peak was determined by fitting a 3rd-order polynomial to the across-time median spectrum and subtracting the fit from the power spectral density plot to remove 1/f noise. From here, the peak could be identified as the most prominent peak in the 7.5 - 14.0 Hz range. Pooling data from all subjects, we determined the median IAF to be 10 Hz.

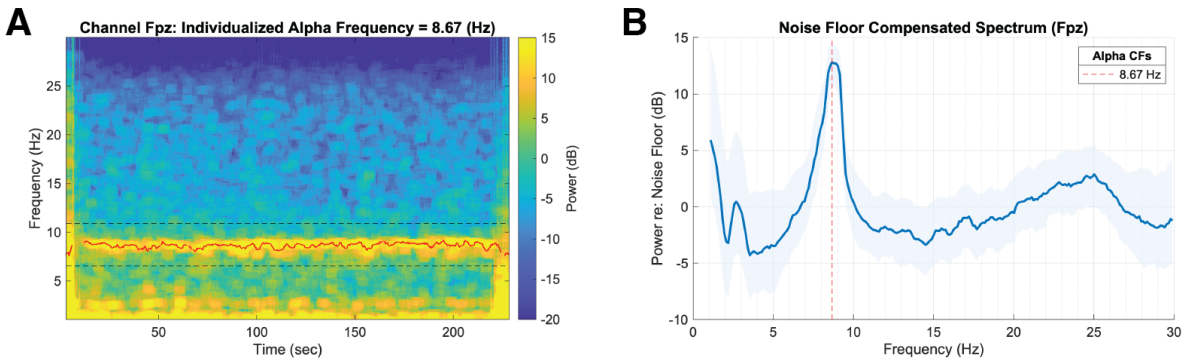

Supplementary Figure 1: Computation of the individual alpha frequency (IAF). A) Multitaper spectrogram computed for one representative subject during a ~2 minute passive recording session. B) Median spectrum over the duration of the recording session shown in A). Red line shows the location of the IAF peak for this subject.

*Estimation of the P1 latency:* To estimate the P1 latency of auditory evoked potentials, we presented 12-ms duration pink noise pulses at approximately 78-82 DB to subjects while they sat with eyes open watching a silent movie with subtitles to pass time. During this time, we presented noise pulses at random phases with at least 750 ms between pulses. Each session lasted approximately 8 minutes, during which approximately 1080 pulses were presented. EEG data was bandpass-filtered between 2 and 30 Hz, and an analysis window was drawn from -250 to 500 ms from the time of sound onset to sort the data into epochs. Any epoch with a peak signal >±100 µV was discarded. The auditory evoked response potential (ERP) was then computed by averaging across epochs. From this average, a P1 search window was defined between 35-75 ms. Pooling data from all subjects, we determined the mean P50 latency to be approximately 62.4 ms.

*Computation of stimulus onset phase to match alpha trough and peak:* Using the computed population average values for IAF and P1, we computed the optimal and pessimal phases according to the formulas below,

○ $Trough\ Phase_{(degrees)} = wrapTo360\left[360^{\circ} \times P1_{(seconds)} \times IAF_{(Hz)}\right]$

○ $Peak\ Phase_{(degrees)} = wrapTo360\left[Trough\ Phase_{(degrees)} - 180^{\circ}\right]$

where $wrapTo360$ is a MATLAB function mapping angles into the range [0, 360].

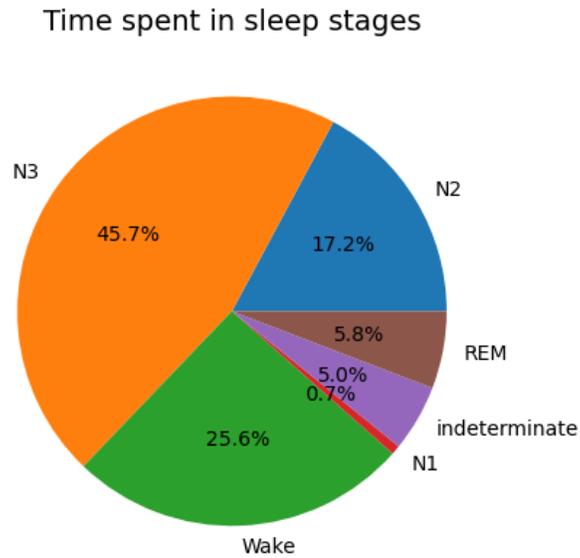

Supplementary Figure 2: Distribution of time spent in each sleep stage for one subject during an overnight recording.

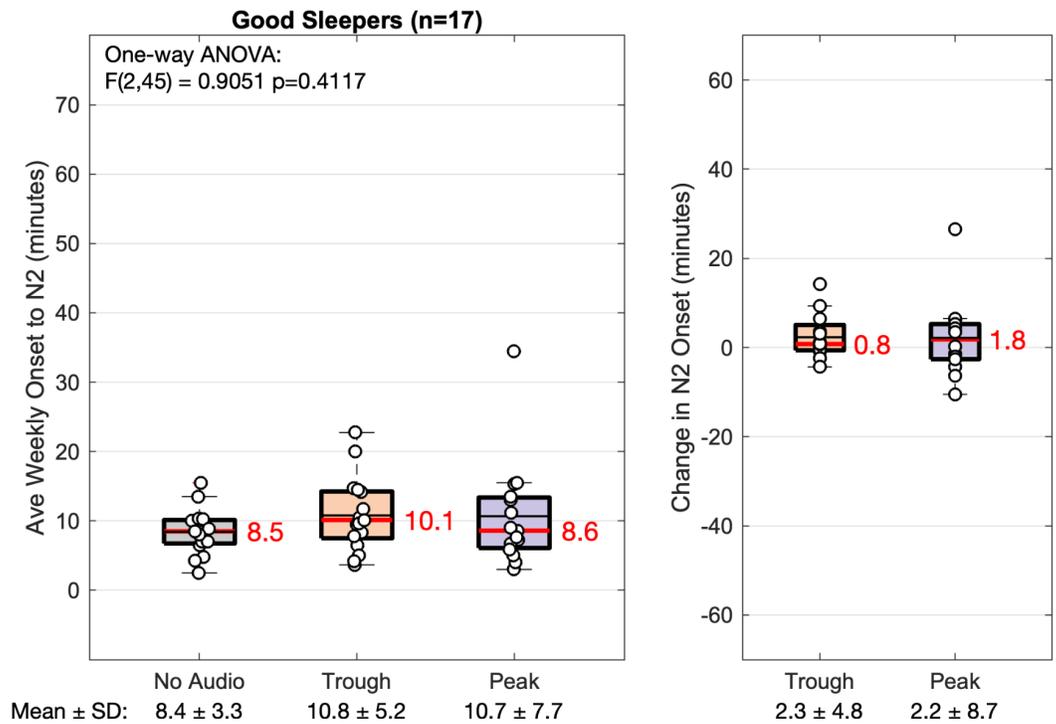

Supplementary Figure 3: Difference in sleep onset latency for individuals who did not have any observable sleep onsets greater than 30 min ("good sleepers"). Box boundaries show the interquartile range, while red lines depict the median. Black lines represent the means of each population. B) Reduction in sleep onset latencies for each stimulation condition relative to control. Markings are as in A.

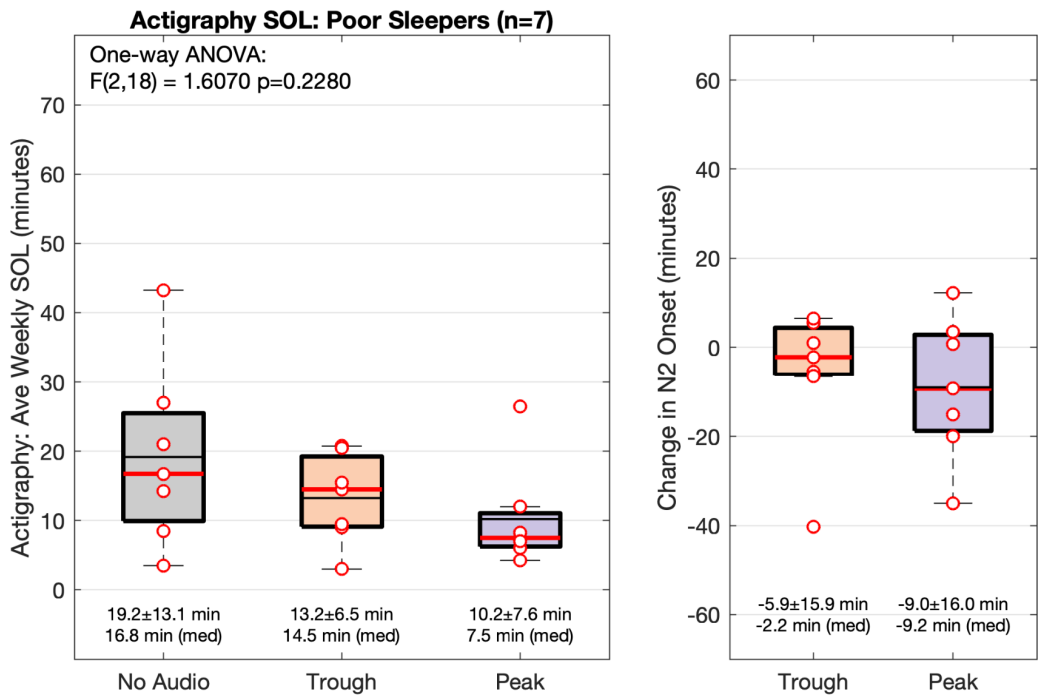

Supplementary Figure 4: Sleep onset latencies for poor sleepers as measured by actigraphy (Philips Actiwatch). Poor sleepers were identified as those subjects whose visually-scored N2 onset latencies were greater than 30 minutes for at least one of the four Control (no audio) nights.

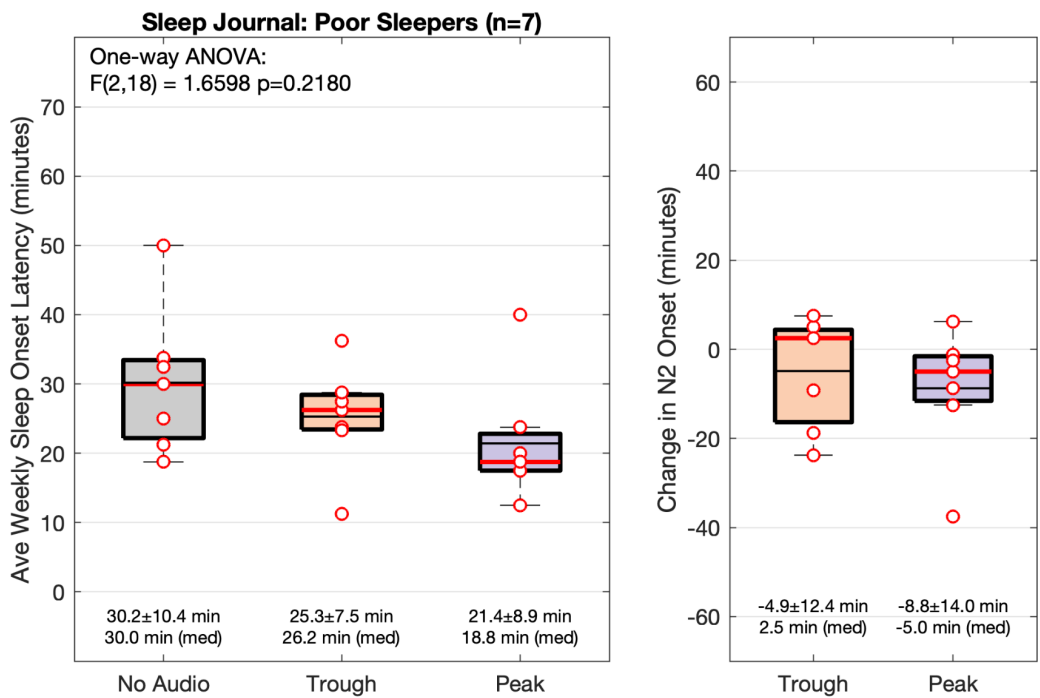

Supplementary Figure 5: Subjective sleep onset latency for poor sleepers as measured by morning survey. Poor

sleepers were identified as those subjects whose visually-scored N2 onset latencies were greater than 30 minutes for at least one of the four Control (no audio) nights.